\documentclass[aps,prd,amssymb,superscriptaddress,nofootinbib,twocolumn,preprintnumbers]{revtex4-1}
\usepackage[utf8]{inputenc}
\usepackage[english]{babel}
\usepackage{amsmath}
\usepackage{amsfonts}
\usepackage{amssymb}
\usepackage{amsthm}
\usepackage{color}
\usepackage{graphicx}
\usepackage[force]{feynmp-auto}
\usepackage{feynmp}
\usepackage[hidelinks]{hyperref}

\newcommand\bw{\begin{widetext}}
\newcommand\ew{\end{widetext}}

\newcommand{\dd}{\,\partial}
\newcommand{\bb}{\,\square}
\newcommand{\nn}{\frac{9\lambda^2}{8\pi^2}}

 \def\be{\begin{equation}}
\def\ee{\end{equation}}
 \def\ba{\begin{align}}
\def\ea{\end{align}}
\def\bea{\begin{eqnarray}}
\def\eea{\end{eqnarray}}

\def\g{\gamma}

\begin{document}

\title{\bf Exact quantum conformal symmetry, its spontaneous breakdown,\\ and gravitational Weyl anomaly}
\author{Mikhail Shaposhnikov}
\email[]{mikhail.shaposhnikov@epfl.ch}
\address{Institute of Physics, 
\'Ecole Polytechnique F\'ed\'erale de Lausanne (EPFL), CH-1015 Lausanne, Switzerland}
\author{Anna Tokareva}
\email[]{a.tokareva@imperial.ac.uk}
\preprint{Imperial/TP/2022/AAT/4}
\address{Theoretical Physics, Blackett Laboratory, Imperial College London, SW7 2AZ London, U.K.}

\begin{abstract}
The classical Lagrangian of the Standard Model enjoys the symmetry of the full conformal group if the mass of the Higgs boson is put to zero.  This is a hint that conformal symmetry may play a fundamental role in the ultimate theory describing Nature. The origin of scales, such as the Higgs vacuum expectation value (vev), may result from the spontaneous breakdown of the conformal symmetry by the dilaton field. In this work, we study whether this classical setup can be implemented in quantum theory and be phenomenologically viable by presenting an explicit construction where the exact conformal symmetry can be preserved and is anomaly free while being spontaneously broken. Not only the Higgs mass but also the genuine quantum scales like the QCD confinement radius are generated by the dilaton vev. We also discuss the extension of these ideas to the theories with dynamical gravity and show that the only finite subgroup of the local Weyl transformations which is anomaly free corresponds to the global scale symmetry. This means that the conformal invariance of the flat space theory is explicitly broken down to the scale symmetry by gravitational effects related to the Weyl anomaly.
\end{abstract}

\maketitle

\section{Introduction}

Can conformal invariance (CI) be an exact symmetry of the fundamental theory of Nature?  Usually, it is thought that this is not the case for three reasons. The first reason is connected to the fact that one of the predictions of the conformal symmetry is the absence of any explicit mass scale in the Lagrangian which is, clearly, not the case: most elementary particles we know have non-zero masses. Second, all realistic {\em renormalisable} field theories, even in the absence of masses, suffer from conformal or scale anomaly: the divergence of the dilatational current is non-zero due to quantum effects and is proportional to the $\beta$-functions of dimensionless coupling constants, governing their renormalisation group running \cite{Coleman:1970je}. And, finally,  the Weyl anomaly \cite{Duff:1977ay,Duff:1993wm,Deser:1993yx}  forbids keeping the classical Weyl symmetry, considered as a generalisation of CI to curved space-time, in a quantum theory in a non-trivial gravitational background metric.\footnote{To evade the confusion:  in what follows, ``scale'' or ``dilatational'' or ``conformal'' anomaly will always refer to the anomaly of the dilatational current in flat-space time whereas ``Weyl'' anomaly will be always associated with non-invariance with respect to Weyl transformations of the effective action in curved space-time.} 

The first problem can be overcome if one introduces a concept of spontaneous breaking of conformal symmetry. If there is a scalar field which can have a non-zero vacuum expectation value all mass scales in the theory can appear without the explicit breaking of the symmetry. As for the second problem -- conformal anomaly -- there are only a few known {\em renormalisable}  theories, all supersymmetric, where the scale anomaly is absent  \cite{Gliozzi:1976qd,Sohnius:1981sn}. An example of non-unitary conformal field theory (CFT) with spontaneous symmetry breaking was constructed in \cite{Karananas:2019fox}. The known conformal anomaly-free  {\em renormalisable}  theories (for example, $N=4$ super-Yang-Mills theory) are far from reality and their relevance for phenomenology is obscure. So, the current answer to the question: ``Do we have a {\em renormalisable field theory}  which is conformally invariant, but CI is spontaneously broken, such that the low energy limit of this theory is the Standard Model?" is ``no''.

What is more important:  renormalisability or conformal symmetry? Though the Standard Model is a renormalisable theory, the ultimate theory of Nature is most probably not. The reason is that the theory of gravity is not renormalisable. Therefore, it makes sense to reformulate the question posed above into the following one: ``Can we have an {\em effective field theory}, valid up to the energies $E_{CI}$ much exceeding the Fermi scale,  which is conformally invariant, but CI is spontaneously broken, such that the low energy limit of this theory is the Standard Model?". The answer to this question is yes \cite{Shaposhnikov:2008xi,Gretsch:2013ooa} (for earlier works see \cite{Wetterich:1987fm, Wetterich:1987fk} and for review  \cite{Wetterich:2019qzx}). 

How can we avoid quantum scale anomaly which causes difficulty in promoting classically conformal invariant theories to the quantum ones? The solution to this problem was suggested almost 50 years ago in \cite{Englert:1976ep}. The reason for the presence of quantum scale anomalies is connected to the fact that any regularisation of divergent Feynman graphs of renormalisable field theories contains an explicit mass scale. It can be a UV cutoff $\Lambda$ or mass $M_{PV}$ in Pauli-Villars regularisation, or the scale $\mu$ in dimensional regularisation (DimReg), eliminating a mismatch between coupling constants in different dimensions. These scales break the conformal invariance explicitly, thus the conformal anomaly. The idea of \cite{Englert:1976ep}, who used  DimReg, consists in replacing $\mu$ by a dynamical field - dilaton $\chi$. This makes the theory conformally invariant in $D=4-2\epsilon$ dimensions and allows the subtraction of divergencies in a conformally-invariant way. The price to pay is the renormalisability of the theory: the Lagrangian in D dimensions contains fractional powers of the dilaton field, leading to the proliferation of different evanescent operators needed to remove the divergencies \cite{Shaposhnikov:2008xi,Shaposhnikov:2009nk}. The spontaneous breaking of the conformal invariance -- the non-zero dilaton vev -- is automatically embedded in the formalism. Of course, the use of DimReg for the construction of CFTs with spontaneously broken CI is not unique, everything works with any type of regularisation: simply replace the cutoff  \cite{Wetterich:1987fm, Wetterich:1987fk}, lattice spacing \cite{Shaposhnikov:2008ar} or the Pauli-Villars mass with the dynamical dilaton field. As usual, the DimReg is more suited to practical computations, which in this context can run up to several loops \cite{Ghilencea:2016ckm}.

Along these lines, the scalar sector of the Standard model can be extended to a theory with spontaneously broken scale invariance in $D$ dimensions by the price of adding an extra scalar field $\chi$ \cite{Shaposhnikov:2008xi},
\begin{equation}
L=\frac{1}{2}(\partial \chi)^2+ \partial H^{\dagger}\partial H-\frac{\lambda}{4}\chi^{\frac{4-D}{D-2}}(H^{\dagger}H-\alpha^2 \chi^2)^2. 
\end{equation}
Here the field $\chi$ has a vacuum expectation value $\chi_0=\langle\chi \rangle=v/\alpha$ (phenomenology requires $\alpha \ll 1$).  The rest of the SM Lagrangian is modified similarly by multiplying the corresponding degree of the dilaton field. The vev of the Higgs field $H$ is $v$. The presence of the field $\chi$ in power proportional to $\epsilon$ generates in the higher loops the higher dimensional operators suppressed by the dilaton vev $\chi_0$ \cite{Shaposhnikov:2008xi,Shaposhnikov:2009nk}, making the effective theory weakly coupled and thus perturbative only below energies $E \lesssim \chi_0=E_{CI}$. In this domain, the predictions of the theory coincide with that of the SM, up to power corrections suppressed by the dilaton vev. The dilaton is an exactly massless particle, being a Goldstone boson of the spontaneously broken conformal symmetry. It only has derivative interactions with matter, if coupled to gravity in a scale-invariant way, see \cite{Wetterich:1987fm,Shaposhnikov:2008xb,Garcia-Bellido:2011kqb,Ferreira:2016kxi} and below.

What happens at energies above $E_{CI}$? If  $E_{CI}$ is the same as the Planck scale $M_P=2.435\times 10^{18}$ GeV, the answer to this question can not be given by the theory in flat space-time, considered so far. Clearly, gravity must be accounted for. If $E_{CI}\ll M_P$, it would be natural to expect that the amplitudes for energies exceeding  $E_{CI}$ should match those of the corresponding CFT without spontaneous symmetry breaking, filling the gap of energies between $E_{CI}$ and $M_P$ (see related discussion in \cite{Karananas:2017mxm}). In this case, no new fields or particles would be required above the naive UV cutoff  $E_{CI}$ to make the theory self-consistent.

In addition to the hopes for consistent UV completion, the theories with exact but spontaneously broken CI are interesting from other points of view. Having conformal invariance of the fundamental theory of Nature may be connected to the explanation of the spin quantisation of massless particles. Indeed, according to the group theory results \cite{Mack:1969dg, Mack:1969rr}, there are no continuous spin representations of conformal symmetry. Although the possibilities of the presence of the massless fields with continuous spin were discussed \cite{Schuster:2014hca}, there are no signs that they exist in nature, though they are admitted by the Poincare group  (a textbook discussion can be found in \cite{Weinberg:1995mt}). Besides the conformal symmetry, there is no strong reason for the absence of these representations. Two other hints are provided by the well-known hierarchies of the Standard Model and gravity: the smallness of the cosmological constant and the Fermi scales in comparison with the Planck scale. Putting the Higgs mass to zero in the classical Lagrangian of the SM leads to enhanced symmetry - the conformal one, leading to a possible explanation of the Higgs mass hierarchy \cite{Wetterich:1983bi,Bardeen:1995kv, Shaposhnikov:2018xkv,Shaposhnikov:2018jag,Shaposhnikov:2020geh}. Moreover, the spontaneous breaking of CI leads to the degeneracy of the vacuum which ensures that the energy of the ground state is equal to zero \cite{Amit:1984ri,Einhorn:1985wp,Rabinovici:1987tf,Shaposhnikov:2008xi}. 
Both these observations are also true if the conformal invariance is replaced by a weaker requirement of the scale invariance.

The discussion above provides us with enough motivation to study further the conformal theories with spontaneous symmetry breaking. To start with, we will concentrate on the theories in flat space-time in various dimensions. 

The first question we are going to elucidate is whether one can indeed remove all divergences in a way consistent with conformal invariance. This problem has been already studied in \cite{Gretsch:2013ooa} with an affirmative answer. The proof given in this paper is based on the assumption that any conformally-invariant operator constructed from a scalar field $\chi$ in flat space-time can be derived from operators invariant under the general transformation of coordinates constructed from the metric $g_{\mu\nu}$ and then reduced to a conformally flat metric by substituting $g_{\mu\nu} \to \chi^2 \eta_{\mu\nu}$. Though this statement is true for non-integer or an odd number of dimensions, it does not hold for even dimensions due to the Weyl anomaly, as we will see below. We provide a simpler (in our opinion) proof which does not require any reference to curved space-time and is based on the background field method (see \cite{Abbott:1980hw} for a review). To clarify even more the absence of conformal anomaly we compute a finite part of the effective action in a toy model and demonstrate that it is indeed conformally invariant within our formalism.

Next, we will construct a most general local effective action for the dilaton, considering it as an expansion over the number of space-time derivatives for a theory defined in an arbitrary number of dimensions. Again, this action is found usually with the help of different curvature invariants in curved space-time (see, e.g. \cite{Baume:2013ika,Osborn:2015rna}). We show how this can be done directly in flat space-time, without any reference to general relativity.

The spontaneous breakdown of the conformal symmetry requires the existence of an exactly flat direction in the effective action for the dilaton, or, what is the same, the absence of the quartic dilaton self-coupling $\propto \chi^4$ in 4-dimensional space-time.  We show that the dilaton flat direction is perturbatively stable with respect to quantum corrections associated with the dilaton field itself. We also discuss a constraint on the dilaton effective action coming from this requirement from certain non-perturbative contributions.

Given that any realistic theory must contain dynamical gravity the question of whether the conformal symmetry in flat space-time can be extended and mapped to some global symmetry which holds in the presence of gravity becomes important.  Since the scale and conformal symmetries do not lead to the conserved charges which can be evaporated from black holes, (contrary to the compact global symmetry groups corresponding, e.g., to the baryon number)  the black hole arguments \cite{Palti:2019pca, Grana:2021zvf} against global symmetries are not applicable to scale or conformal invariance.

A natural extension of the conformal invariance from the flat space-time to curved space-time is the Weyl symmetry -- invariance of the theory against replacing of the metric  $g_{\mu\nu} \to \Omega^2(x) \g_{\mu\nu}$, where $ \Omega(x)$ is an arbitrary function of space-time, matched by the corresponding transformation of the matter fields  (see, e.g. \cite{Komargodski:2011vj,Luty:2012ww}). It is well known that this symmetry is anomalous  \cite{Duff:1977ay,Duff:1993wm,Deser:1993yx} and thus cannot be kept at the quantum level.  We discuss the reasons why this happens and address the question of what is the maximal global subgroup of the local Weyl group that can be anomaly free in arbitrary gravitational background.  We find that among the Weyl transformations, only dilatations can stay as an exact quantum symmetry. We derive a general condition for the selection of the metrics which admit a subgroup of the Weyl symmetry related to the conformal symmetry of flat space-time, the flat space-time obviously satisfies this condition.

The paper is organised as follows. In the next Section, we review the basic notions and the properties of the conformal group needed for further discussions. Section III is devoted to a method allowing keeping the conformal symmetry in all orders of perturbation theory. We also give here an explicit computation of the effective action in a simple toy model demonstrating its CI. In Section IV we derive the dilaton effective action in an arbitrary dimension of space-time. In Section V we discuss the method for the construction of the dilaton action with the use of the curvature invariants. In Section VI we elucidate the origin of the Weyl anomaly. In Section VII we discuss several subgroups of the local Weyl group and demonstrate that only scale symmetry can survive the quantum anomaly. In Section VIII we discuss a generic scale-invariant Lagrangian for the dilaton and gravity. In IX we conclude.

A short account of the obtained results is presented in \cite{Shaposhnikov:2022dou}. Similar questions were also studied in \cite{Hinterbichler:2022ids}.

\section{Preliminaries}
\label{sec:preliminaries}

Here we review the basic grounds of the conformal transformations. Let $\Phi$ be a multicomponent field\footnote{This field can have both Lorentz and internal indices which correspond to the transformations in the field space. For simplicity, we omit these indices.} transforming conventionally under Poincare transformations -- translations $\delta_T$ and rotations $\delta_L$ -- which leave the action invariant,
\begin{equation}
\delta_T^{\sigma}\Phi(x)=\dd_{\sigma} \,\Phi(x),
\end{equation}
\begin{equation}
\delta_L^{\sigma\tau}\Phi(x)=(x^{\sigma}\dd^{\tau}-x^{\tau}\dd^{\sigma}+\Sigma^{\sigma\tau} )\,\Phi(x).
\end{equation}
Here $\Sigma^{\sigma\tau}$ is spin matrix (SO$(D-1,1)$) acting on a given representation of the field $\Phi$, $D$ is the dimensionality of space-time.

In addition to these transformations, we consider dilatations 
\begin{equation}
\delta_S \Phi(x)=(x_{\tau}\dd^{\tau}+	\Delta) \Phi(x),
\end{equation}
and special conformal transformations,
\begin{equation}
\delta_C^{\sigma} \Phi(x)=(2 x^{\sigma}x^{\tau}-\eta^{\sigma\tau} x^2)\dd_{\tau}\Phi(x)+2x_{\tau}(\eta^{\sigma\tau}\Delta-\Sigma^{\sigma\tau})\Phi(x).
\end{equation}
Here $\Delta$ is a mass dimension of the field. For the canonically normalised scalar fields $\Delta=(D-2)/2$ in $D$-dimensional spacetime. 

All listed transformations preserve the angles between any lines in the space. They can be also defined as the coordinate transformations which lead to the rescaling of the flat spacetime metric by some function of coordinates,
\begin{equation}
\label{conformal-def}
    x'_{\mu}=F^{\mu}(x),\quad \eta'_{\mu\nu}=\eta_{\lambda\sigma}\frac{\partial F^{\lambda}}{\partial x_{\mu}}\frac{\partial F^{\sigma}}{\partial x_{\nu}}=\Omega^2(x)\eta_{\mu\nu} ~ .
\end{equation}
All the solutions to this requirement form a group SO$(D,2)$ with $(D+ 1)(D+ 2)/2$ generators. Poincare transformations are described by the ISO$(D-1,1)$ subgroup with $D(D+ 1)/2$ generators leaving the spacetime intervals invariant. The dilatations rescale the coordinates by a constant factor, $x_{\mu}'=\lambda x_{\mu}$ while the finite special conformal transformation parametrised by the constant vector $c_{\mu}$ looks,
\begin{equation}
\label{x-trasform}
x_{\mu}\rightarrow x'_{\mu}=\frac{x_{\mu}+c_{\mu}x^2}{\Omega(x,c)},
\end{equation}
where $\Omega(x_{\mu},c_{\mu})=1+2 c_{\mu}x_{\mu}+c^2 x^2$. 
The corresponding finite transformation of the field $\Phi$ depends on the spin. For the simplest case of the scalar, the transformation is,
\begin{equation}
\label{ctphi}
\Phi'(x')=[\Omega(x,c)]^{\frac{D-2}{2}}\Phi(x).
\end{equation}
These coordinate transformations make the flat space metric $\eta_{\mu\nu}$ transformed to
\begin{equation}
\label{flatct}
   \eta_{\mu\nu}\rightarrow \Omega(x,c)^{2}\eta_{\mu\nu}.
\end{equation}

Special conformal transformations can be written as a combination of inversion,
\begin{equation}
x'_{\mu}=C x_{\mu}/x^2,
\end{equation}
shift and rotation. This means that, for practical applications, it is enough to prove the invariance of the Lagrangian under consideration with respect to the inversion only. This invariance, together with the scale symmetry, would mean that the action has the full conformal symmetry SO$(D,2)$.

The infinitesimal conformal transformations $x_{\mu}'=x_{\mu}+\xi_{\mu}$ can be also defined through the equation on the Killing vector $\xi_{\mu}$ following from the definition \eqref{conformal-def},
\begin{equation}
\label{killing}
\partial_{\mu}\xi_{\nu}+\partial_{\nu}\xi_{\mu}=\frac{2}{D}\eta_{\mu\nu}\,\partial_{\alpha}\xi^{\alpha}.
\end{equation}
In this case, the metric gets rescaled by the factor $\Omega^2=1+\omega=1+\partial_{\alpha}\xi^{\alpha}$. Differentiating \eqref{killing} twice one can get a closed form equation for $\omega$,
\begin{equation}
\label{sigma_constraint}
   (2-D) \partial_{\mu}\partial_{\nu}\omega=\eta_{\mu\nu}\bb\omega.
\end{equation}
This equation has only linear in $x_{\mu}$ solution, reflecting the fact that transformations of the conformal group lead to a change of the metric by a conformal factor $\omega$ constrained by the condition \eqref{sigma_constraint}. Remarkably, all the solutions to these equations on $\omega$ correspond to the spacetime conformal transformations. Thus, the transformations can be described by one scalar function. This definition allows extending the conformal group to the case of the curved space which will be discussed in detail later in Sec. 7.

Notice that in $D\ne 2$ this equation reduces to
\begin{equation}
    \partial_{\mu}\partial_{\nu}\omega=0,
\end{equation}
while in $D=2$ one can get less constrained $\sigma$, satisfying only the condition 
\begin{equation}
    \bb \omega=0.
\end{equation}
This condition has more solutions reflecting the fact that the conformal group in $D=2$ has an infinite number of generators (see for example the review \cite{Rychkov:2016iqz} for more details).

In what follows we will be also working with the extensions of the given definitions to the case of the spacetimes with fractional dimensions. All the expressions with fractional values of $D$ appearing while using dimensional regularization of the Lagrangian formulations should be understood as an analytic continuation of the $D$-dependence in the given integer $D$ definitions. Let us mention here that we will use the fact that in the Lagrangian formulation of the CFTs there are no operators whose conformal transformation would be singular around any integer $D$. We will use this fact in our arguments justifying the existence of anomaly-free quantum conformal symmetry in flat spacetime.

\section{Effective CFTs with the spontaneous breaking of conformal symmetry}
\label{sec:effCFT}
There was a long discussion in the literature regarding the question of whether the breaking of conformal symmetry by quantum anomaly can be avoided in quantum theory. Even a less constraining symmetry -- dilatations -- cannot be kept at the quantum level in a {\em conventional approach} to defining the QFT. The deep reason behind the appearance of the scale anomaly is connected to the fact that the standard multiplicative renormalisation procedure of divergent Feynman diagrams is not compatible with preserving scale symmetry. For example, the Pauli-Villars regularisation requires introducing a large energy scale $M_{PV}$ (eventually sent to infinity) serving as a mass of auxiliary fields which breaks the symmetry.  The dimensional regularisation also cannot save the scale symmetry because it requires the presence of dimensionful parameter $\mu$ needed to match the mass dimensions of couplings in the spaces with different dimensions. In a classical conformal or scale-invariant field theory, one can define a dilatational current, which is conserved on equations of motion. In quantum theory, the explicit breaking of the scale symmetry by regulators leads to scale anomaly \cite{Coleman:1970je} - the divergence of the dilatational current is not zero any longer and is proportional to $\beta$-functions describing the renormalisation group running of different coupling constants. In the words of Coleman \cite{Coleman:1985rnk}: ``For scale invariance, though, the situation is hopeless; any cutoff procedure necessarily involves a large mass, and a large mass necessarily breaks scale invariance in a large way.''

However, there is a simple way out of this assertion, going back to \cite{Englert:1976ep} (see also \cite{Wetterich:1987fm}).  Make the cutoff (or Pauli-Villars mass, or parameter $\mu$ of DimReg) dynamic and proportional to one of the fields in the theory, say the ``dilaton'' $\chi$. Arrange the theory in such a way that $\chi$ has a vacuum expectation value (vev). Then the theory is automatically scale-invariant in all orders of perturbation theory, there is no anomaly in the dilatational current, but the scale symmetry is ``hidden'', i.e. spontaneously broken. This procedure breaks the renormalizability of the theory - it generates an infinite number of higher-dimensional operators different from those in the original classical action. These operators are suppressed by the dilaton vev, which can be as large as the Planck scale. The theory resulting from this construction is a well-defined and predictive effective field theory up to the energy scale $\sim \langle \chi \rangle$. One may expect that at higher energies it maps to an unbroken conformal theory. A generic feature of a theory with spontaneously broken scale invariance is the presence of an exactly massless Goldstone particle - the dilaton. In a theory without dynamical gravity, this massless particle generically induces a long-range (fifth) force, which disappears if gravity is added in a scale-invariant way \cite{Wetterich:1987fm,Shaposhnikov:2008xi,Ferreira:2016kxi}. The scale-invariant Standard Model defined in this way was widely discussed in phenomenological and cosmological contexts \cite{Shaposhnikov:2008xi,Garcia-Bellido:2011kqb}.

The scale invariance of the effective field theory constructed along these lines is obvious - there are simply no dimensionful coupling constants or explicit mass parameters in it. It is far from being obvious, though, that the effective action can be invariant under the full conformal group, including the special conformal transformations. This question is important since the number of conformally invariant operators is smaller than the number of scale-invariant ones (see Section \ref{sec:effact}).

In a nice article, \cite{Gretsch:2013ooa},  it was claimed that this is indeed the case in all orders of perturbation theory for theories without the gravitational anomaly (i.e. invariant under all diffeomorphisms (Diffs) at the quantum level). The first step of the proof was to couple the original conformally invariant theory in D dimensions (D is fractional in DimReg, $D=4-2\epsilon$) to the external background metric in a Weyl-invariant way. Then the one-loop regularised (but not renormalised yet) effective action enjoys both Diff and Weyl invariance. With the use of the Diff invariance, the authors argued that the pole part of this expression $\propto 1/\epsilon$ can be made conformally invariant, meaning that the theory indeed remains conformal at the one-loop level. Now, one can proceed with an iterative procedure: add this one-loop counter-term to the classical action, repeat the computation to build the two-loop conformally invariant counter-term, etc, etc.

It looks to us, however, that this proof is not fully satisfactory because it calls for getting out from the flat space and exploits Weyl symmetry which is anomalous \cite{Duff:1977ay,Duff:1993wm,Deser:1993yx}. Indeed, it is known \cite{Karananas:2015ioa} that not all conformal operators can be written in a curved space as Weyl invariant operators in a specific spacetime dimension without poles in $\epsilon$ coefficients (see also Section \ref{sec:anomaly}). Potentially, these extra singularities may jeopardise the flaw of logic separating the $1/\epsilon$ infinities associated with renormalisation from $1/\epsilon$ terms associated with the Weyl anomaly.

For this reason, we present another argument that the perturbative effective action has conformal symmetry if the theory is regularised in a conformally-invariant way. It is a straightforward application of the background field method \cite{Abbott:1980hw} and it does not require getting out of flat space.

To clarify the main ideas we start with a one-loop computation in a toy model which shows how the conformal symmetry is kept. Then we will turn to a general case.

\subsection{Effective action and its conformal invariance in a toy model}

In this section, we examine an explicit example of one-loop effective action for a conformal scalar field.

Consider the action defined in $D$ dimensions,
\begin{equation}\label{l4}
{\cal L}=\frac{1}{2}(\partial\phi)^2-\frac{\lambda}{4}\phi^{q}.
\end{equation}
Here $q=2 D/(D-2)$. Let us compute the simplest one-loop graph in the background field method \cite{Abbott:1980hw}. To apply this method we expand the field as $\phi=\phi_0+\delta \phi$. Then, the perturbed Lagrangian becomes,
\begin{equation}\label{eq:2order}
{\cal L}=\frac{1}{2}(\partial (\delta\phi))^2-\frac{\lambda}{4} \frac{q(q-1)}{2}\phi^{q-2}\delta \phi^2.
\end{equation}
\vspace{20pt}

\begin{figure}[htb]
    \centering
    \includegraphics{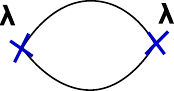}
    \caption{The leading contribution to the effective action in a background field formalism for the model \eqref{l4}.}
    \label{fig:my_label}
\end{figure}
The propagator for massless field $\delta \phi$ is
\begin{equation}
\langle \delta \phi(x)\delta \phi(y)\rangle=\left(\frac{1}{(x-y)^2}\right)^{\frac{D-2}{2}}.
\end{equation}
In the formalism of effective action, the background-dependent mass term in \eqref{eq:2order} is treated as an interaction term since we use the perturbative expansion of the background itself. For this reason, we can use the massless propagator in further computation. The effective action at the order $\lambda^2$ is described by the diagram shown on Figure 1,
\begin{equation}
\Gamma=\nn\int d^D x d^D y\phi(x)^{\frac{4}{D-2}}\phi(y)^{\frac{4}{D-2}}\left(\frac{1}{(x-y)^2}\right)^{D-2}~.
\end{equation}
The propagator can be written in Fourier space making the divergence of the effective action in 4 dimensions explicit,
\bw
\begin{equation}
    \left(\frac{1}{(x-y)^2}\right)^{D-2}=-\frac{1}{(2\pi)^D}\frac{\pi^2}{D-4}\int\left(p^2\right)^{\frac{D-4}{2}} e^{ip(x-y)}d^D p+ {\rm (finite)}~.
\end{equation}
\ew

The counterterm which is conformal in $D$ dimensions has the form,
\begin{equation}
{\cal L}_{ct}=\frac{9}{8\pi^2}\frac{1}{D-4}\lambda^2 \int d^D x\, \phi^{\frac{2 D}{D-2}}(x).
\end{equation}
Let us now compute the renormalized effective action,
\begin{equation}
\begin{split}
\Gamma_{ren}=\nn \int d^D p\, e^{ip(x-y)}\int d^D x d^D y \\ \left(\phi^{\frac{D}{D-2}}(x)\phi^{\frac{D}{D-2}}(y)-\phi^{\frac{4}{D-2}}(x)\phi^{\frac{4}{D-2}}(y)(p^2)^{\frac{D-4}{2}}\right).
\end{split}
\end{equation}
The limit of this expression for $D\rightarrow 4$ is finite,
\begin{equation}
\begin{split}
\Gamma_{ren}=\nn \int d^4 p e^{ip(x-y)}\int d^4 x d^4 y \\ \phi^2(x)\phi^2(y)\log{\frac{\phi(x)\phi(y)}{p^2}}.
\end{split}
\end{equation}
This non-local effective action can be rewritten in the form of only one $x$ integral with a non-local differential operator,
\begin{equation}
(\log\bb)\,\phi(x)=\int d^4 y K(x-y)\phi(y)
\end{equation}
defined through its kernel
\begin{equation}
K(x-y)=\int d^4 p \log{(-p^2)} e^{ip(x-y)}
\end{equation}
as follows
\begin{equation}
\label{Gamma_eff}
\begin{split}
&\Gamma_{ren}=\nn\left( \int d^4 x \,\phi^4(x)\log{\frac{\phi^2(x)}{\mu^2}}-\right. \\ &\left.- \int d^4 x d^4 y d^4 p\,\phi^2(y)\phi^2(x)\log{\frac{p^2}{\mu^2}}\right) = \\ =& \nn\left( \int d^4 x\, \phi^4(x)\log{\frac{\phi^2(x)}{\mu^2}}\right.- \\ & -\left.\int d^4 x\,\phi^2(x)\log{\frac{-\bb}{\mu^2}}\phi^2(x)\right)=\\&=-\nn \int d^4 x\,\phi^2(x)\log{\frac{-\bb}{\phi^2(x)}}\phi^2(x).
\end{split}
\end{equation}

The conformal invariance of this operator is not obvious. To see that, let us extend the validity of the formula (\ref{boxes}) derived in Section \ref{sec:effact} to the values of $N$ which are not integer\footnote{See also \cite{Manvelyan:2007tk,Joung:2015jza} for the discussion of the fractional powers of the box operator.}. This can be defined for example as an analytic continuation of the result known for integer numbers to the whole complex plane. Notice that in 4 dimensions, as follows from (\ref{boxes}), $\phi^{2-N}\bb^N \phi^{2-N}$ is conformally invariant at any $N$. In the limit $N\rightarrow 0$ we can write formally,
\begin{equation}
\phi^{2-N}\bb^N \phi^{2-N}=\phi^4+N\phi^2\left(\log{\frac{\bb}{\phi^2}}\right)\phi^2+\dots
\end{equation}
Then, the operator in (\ref{Gamma_eff}) can be obtained as,
\begin{equation}
\label{limit}
\phi^2\left(\log{\frac{\bb}{\phi^2}}\right)\phi^2=\lim_{N\to 0} \left(\frac{\phi^{2-N}\bb^N\phi^{2-N} -\phi^4}{N}\right).
\end{equation}
In Section \ref{sec:effact} we will construct the conformal operators out of the fields and $\bb$-operators. As a specific consequence of this result, the operator $\bb^N\phi^{2-N}$ is conformally-covariant for all $N$ in four dimensions (see also \cite{Manvelyan:2007tk,Joung:2015jza}). In the spirit of analytic continuation from all integer $N$ to all real powers, the limit $N\rightarrow 0$ of \eqref{limit} appears to be conformal, hence the effective action $\Gamma_{ren}$ remains conformal.

Our result on conformal invariance of the one-loop effective action for the theories defined as conformal in $D$ dimensions can be generalized to the next orders in perturbation theory. Since both effective action and counter-term have this symmetry in $D$ dimensions, the renormalized effective action will be invariant in 4 dimensions, too. This conclusion would be obvious if the transformations do not depend on the number of dimensions. However, the same result holds also for our case of conformal symmetry for the following reason. Conformal transformations depend on $D$ in such a way that they do not bring any pole parts proportional to $1/(D-4)$, see Section 2 for the definitions. The result of the conformal transformation of the regular in $D=4$ terms in the action remains regular. For this reason, there are no extra non-invariant terms left after renormalisation. 

\subsection{General consideration}

For concreteness, let us consider a theory invariant under the conformal transformations in $D=4$, though the results are valid for any integer dimension.  With the help of the dilaton field raised into the fractional power, every CI term in the Lagrangian can be extended in a CI way to the fractional number of dimensions. Hereafter we will use dimensional regularisation allowing us to define the theory as conformal in $4-\epsilon$ dimensions. Our observation can be formulated as follows:

{\it
If the theory is defined in such a way that
\begin{itemize}
    \item the action is invariant under the conformal transformations in an arbitrary (also fractional) spacetime dimension, \footnote{We define the theory in a fractional dimension as an analytic continuation of its results in an integer dimension}
    \item the dimensional regularization is used for computations of divergent terms in the effective action in an integer number $D$ of dimensions,
    \item and  the perturbative expansion can be applied,
\end{itemize}
then the finite renormalized effective action $\Gamma_{ren}$ is invariant under the conformal transformations in $4$ dimensions.
}

Let us start the argumentation from the one-loop level. The computation of the effective action would lead to the appearance of the divergences which need to be regularized. Since all the procedure of the computations explicitly keeps the conformal invariance in any dimension the result would be invariant under the conformal transformations in $4-\epsilon$ dimensions. However, the result will be divergent in $4$ dimensions. The divergent part of the total effective action $\Gamma_{div}+\Gamma_{fin}$ is,
\begin{equation}
    \Gamma_{div}=\frac{\Gamma_1}{\epsilon}+\dots~.
\end{equation}
Thus, we have
\begin{equation}
 \delta_{D-\epsilon}(\Gamma_{div}+\Gamma_{fin})=0~.
\end{equation}
 
To cancel these divergences one has to introduce counterterms $L_c$. They can be written in a form of the local operators invariant under the conformal transformations in $4-\epsilon$ dimensions. This can be done because the number of conformal operators in $4-\epsilon$ dimensions is equal or larger than in exactly $4$ dimensions, given that in integer dimensions some operators can reduce to total derivatives. Let us stress here that the counterterms should be extended to the conformal operators in $4-\epsilon$ by adding only finite at $\epsilon\rightarrow 0$ pieces\footnote{This would not be possible for the Weyl transformations, since some operators get accompanied by extra $1/\epsilon$ contributions.}. This can be done for the conformal transformations, so $L_c$ can be chosen such that
\begin{equation}
    \delta_{D-\epsilon}L_c=0~.
\end{equation}
Let us compute the transformation of the finite part left after the renormalization,
\begin{equation}
    \delta_{D-\epsilon}\Gamma_{ren}=\delta_{D-\epsilon}\Gamma_{div}+\delta_{D-\epsilon}\Gamma_{fin}+\delta_{D-\epsilon} L_c=0~.
\end{equation}
Using essential facts about the conformal transformations, namely,
\begin{itemize}
 \item $ \delta_{D-\epsilon}{\cal O}-\delta_{D}{\cal O}=O(\epsilon)B$, where operators ${\cal O}$ and $B$ are finite at $\epsilon\rightarrow 0$,
\item $\delta_{D-\epsilon}{\cal O}~{\rm is~ finite~at~}\epsilon\rightarrow 0$,
\end{itemize}
we can conclude that the renormalized effective action at one loop level stays invariant under the conformal transformations in $D$ dimensions. Notice also that from these properties it follows that $\Gamma_{div}$ and $\Gamma_1$ are invariant under the $D$-dimensional conformal transformations separately.

Going to the two-loop level, we have to add the one-loop counterterm and repeat the procedure. The new issue arising at the two-loop level is the presence of different types of divergences in the regularized effective action: besides $1/\epsilon$ there will be $1/\epsilon^2$ terms. In a full analogy with the one-loop case, we can build a conformal in $4-\epsilon$ dimension counterterm which cancels $1/\epsilon^2$ divergence. However, in general, this counterterm will have a $1/\epsilon$ part itself. The key observation here is the fact that $1/\epsilon$ term will be invariant with respect to $4$-dimensional conformal transformations, in the same way, as $\Gamma_{fin}$ was invariant at a one-loop level. As the operator is conformal in $4$ dimensions, it can be extended to $4-\epsilon$ dimensions in a conformally invariant way (see the explicit construction \eqref{conf_O}). Thus, one can build proper counterterms cancelling $1/\epsilon$ divergences arising from two loops. The finite part of the effective action will be kept conformal. 
Iteratively, the described procedure can be repeated for an arbitrary number of loops. This way, the conformal invariance of the renormalized effective action can be justified step by step in a perturbative expansion.

In the same way, one can investigate whether any other symmetry of the classical action will be preserved at the quantum level in perturbation theory. The key point here is that at each step of the consideration no extra divergences are appearing during our attempt to make the counterterm invariant. For example, there are no $1/\epsilon^2$ terms required to make one-loop counterterm conformal in $D$ dimensions. As we will see later in Sec. \ref{sec:anomaly}, it is precisely this condition which is not satisfied for Weyl symmetry which causes Weyl anomaly. 

\section{Dilaton effective action}
\label{sec:effact}

The low energy limit on any spontaneously broken conformal field theory contains a massless scalar - dilaton - the Goldstone boson. The realistic theory would also contain a massless photon. The addition of the massless vector field can be made without difficulties and thus is not considered in what follows. This section aims to construct all possible local conformally invariant operators contributing to the dilaton effective action. We classify them by the number of derivatives. The generic effective Lagrangian would contain all the operators respecting the conformal symmetry, and can be used for perturbative computations at low momentum transfers.

Let us take the flat space-time of dimensionality D, where D can be fractional. It is easy to construct scale-invariant operators. Simply write all Poincare invariant operators containing different powers of the scalar field and its derivatives, and make sure that the operator has a mass dimension D.  Not all scale-invariant operators are conformally invariant. For instance, the operator $\phi^{2D/(2-D)}(\dd \phi)^4$ is scale- but not conformally invariant. 

To find a subclass of conformal operators, one can proceed as follows. Any scale-invariant operator, after several integrations by parts in the action, can be written as a product of a scalar field in some (possibly fractional) power and boxes $\square$ (D'Alambertians)  in integer powers (remember, we only consider local operators, differential operators in fractional powers would introduce non-locality) acting on some power of the scalar field. As the scalar field transforms uniformly under the conformal transformation (ref{ctphi}), to find the conformal operators it is sufficient to determine the power $\alpha$ in the expression $\square^N \phi^\alpha$, where $N$ is an integer number,  making it to transform homogeneously under CT.  

We can use the fact that each transformation from the conformal group can be presented as a combination of translations, rotations, dilatation and inversion, see Section \ref{sec:preliminaries}. 
Note that the rotations, translations and dilatations can lead only to the rescaling of $\square^N \phi^\alpha$ by a constant factor. So, the only non-trivial part here is the inversion which in Euclidean spherical coordinates means a change of the radial coordinate $r \to 1/r$, while the angles remain unchanged. The respective transformation of the canonically normalised field $\phi$ is
\begin{equation}
\label{inver}
\phi(r) \to \tilde\phi(r) \equiv r^{(2-D)} \phi\left(\frac{1}{r}\right)~.
\end{equation}
Thus, the question boils down to the following. Suppose that $\phi^\alpha(r)$ is a solution to equation 
\begin{equation}
\label{eqbox}
\Box_{r}^N\phi^\alpha(r)=0~,
\end{equation}
where $\Box_{r}$  is the radial part of the Laplace operator in $D$ dimensions
\begin{equation}
\Box_{r}=\frac{\partial^2}{\partial r^2}+\frac{D-1}{r}\frac{\partial}{\partial r}~.
\end{equation}
Can we find the power $\alpha$ in such a way that the field $\tilde \phi^\alpha$ also satisfies the equation $\Box_{r}^N\tilde\phi^\alpha(r)=0$?

The answer to this question can be found easily with the use of Wolfram's Mathematica and reads
\begin{equation}
\alpha=\frac{D-2N}{D-2}.
\end{equation}
Therefore, the operators which transform uniformly under conformal transformations and have the same scaling dimension as $\phi$ can be written as
\begin{equation}
\label{boxes}
O_N=\phi^{\frac{2(N+1)}{2-D}}\square^N (\phi^{\frac{D-2N}{D-2}})~.
\end{equation}

With this basic block, the most general action can be constructed from these operators using multiplications and compositions. If we define
\begin{equation}
\label{ONhat}
\hat{O}_N=\phi^{\frac{2(N+1)}{2-D}}\bb^N,
\end{equation}
then all possible combinations of the form 
\begin{equation}
\label{conf_O}
\hat{O}_{N_1}(O_{m_1}...O_{m_k})\hat{O}_{N_2}(O_{n_1}...O_{n_i})...\hat{O}_{N_p}(O_{s_1}...O_{s_l})
\end{equation}
multiplied by a proper power of $\phi$ (which leads to the scale invariance) would be conformally invariant operators.  To the best of our knowledge expression (\ref{conf_O}) is new and has not appeared in the literature.

Let us notice here the following important observations. In $D\ne 2$ the construction of conformal operators \eqref{conf_O} is done in such a way that 
\begin{itemize}
    \item they can be written for an arbitrary number of dimensions, including fractional $D$ which is needed in DimReg,
    \item there are no singularities in these expressions in an integer number of dimensions. However, some operators or their combinations may become a total derivative in a certain number of dimensions.
\end{itemize}

\subsection{Derivative expansion}
Let us illustrate the obtained result for a few operators with an increasing number of derivatives.
\begin{enumerate}
\item No derivatives:
\begin{equation}
\label{0der}
O_0=\phi,\quad L_0\propto \phi^{2 D/(D-2)}~.
\end{equation}
\item Two derivatives:
\begin{equation}
\label{2der}
O_1=\phi^{\frac{4}{2-D}}\bb \phi, \quad L_2=\phi\bb\phi~.
\end{equation}
\item Four derivatives:\\
 There are three operators, namely  $O_1^2$,  $\hat{O}_1 O_1$ and
\begin{equation}
O_2=\phi^{\frac{6}{2-D}}\bb^2\phi^{\frac{D-4}{D-2}}~.
\end{equation}
Thus, the conformal Lagrangian containing 4 derivatives can be written in the form,
\begin{equation}
\label{4der}
\begin{split}
L_4=a_1 \phi^{\frac{D+2}{D-2}} O_2+a_2 \phi^{\frac{2}{D-2}} O_1^2+a_3 \phi^{\frac{D+2}{D-2}} \hat{O}_1 O_1\\  =  \bar{a}_1 \phi^{\frac{D-4}{D-2}}\bb^2\phi^{\frac{D-4}{D-2}}+\bar{a}_2 \phi^{\frac{4}{2-D}}(\bb\phi)^2.
\end{split}
\end{equation}
Notice that the last term in the first line is equivalent to  $O_1^2$ after omitting the total derivative.
\item Six derivatives.

The conformal operators are constructed as combinations of $O_1$, $O_2$ and $O_3=\phi^{\frac{8}{2-D}}\bb^3(\phi^{\frac{D-6}{D-2}})$. Although there are many combinations, ($O_1^3,~O_1 O_2,~O_3,~\hat{O}_1(O_2),~\hat{O}_1(O_1^2)$,...), only 4 of them are independent and not connected by adding total derivative. One can choose for example
\begin{equation}
\begin{split}
     L_6=& b_1\phi^{\frac{D+2}{D-2}} O_3+b_2 \phi^{\frac{4}{D-2}} O_1 O_2+\\ &+b_3\phi^{\frac{6-D}{D-2}} O_1^3+b_4 \phi^{\frac{4}{D-2}} O_1 \hat{O}_1 (O_1).
    \end{split}
\end{equation}
The Lagrangian then takes the form
\begin{equation}
\label{6der}
\begin{split}
    L_6= & b_1\phi^{\frac{D-6}{D-2}}\, \square^3\left(\phi^\frac{D-6}{D-2}\right)+\\
     & +b_2 \phi^{-\frac{6}{D-2}}\, \square^2\left(\phi^\frac{D-4}{D-2}\right)\square\phi+\\ &+b_3\phi^{-\frac{D+6}{D-2}}\, (\square \phi)^3+b_4 \phi^{-\frac{4}{D-2}} \square\phi\square\left(\phi^{\frac{4}{2-D}}\,\square\phi\right).
    \end{split}
\end{equation}
\end{enumerate}

One can see that all the expressions (\ref{0der}-\ref{6der}) contain singularities at $D=2$, whereas formulas (\ref{4der}) and (\ref{6der}) are singular at $D=4$ and $D=6$ respectively. This happens to be related to the Weyl anomaly which will be discussed in detail in Sec.\ref{sec:anomaly}. In the next subsections, we present the explicit formulas for conformally invariant operators in 4, 6 and 2 dimensions removing these singularities.

\subsection{Operators in $D=4$}
Let us work now in the space with an integer number of dimensions $D=4$. The construction of conformally invariant operators can be done straightforwardly - first, consider the scale-invariant operators and then select only those which are invariant under infinitesimal special conformal transformations. The operators with zero or two derivatives are simply the self-interaction $\lambda \phi^4$ and the kinetic term $ \phi\bb\phi$. All scale-invariant operators containing 4 derivatives can be written through the following set of three operators,
\begin{equation}
\label{4d_all}
\frac{1}{\phi^2}(\bb\phi)^2,~~\frac{1}{\phi^3}(\dd_{\mu}\phi)^2\bb\phi,~~\frac{1}{\phi^4}(\dd_{\mu}\phi)^4.
\end{equation}
All other operators can be recast in terms of these up to the total derivatives.  The requirement of invariance of the action against infinitesimal special transformations shows that only the following two operators are admitted,
\begin{equation}
\label{4d conformal}
\frac{1}{\phi^4}(\dd_{\mu}\phi)^4-\frac{2}{\phi^3}(\dd_{\mu}\phi)^2\bb\phi,~~\frac{1}{\phi^2}(\bb\phi)^2.
\end{equation}
The same result can be derived with the use of the formula (\ref{4der}) of the previous subsection.  The first operator at the second line in (\ref{4der})  looks like the trivial zero contribution in $D=4$. However, we can perform the expansion around $D=4$ with the leading term
\begin{equation}
\left((D-4)\bb\,\log{\frac{\phi}{\mu}}\right)^2.
\end{equation}
Because of the derivative acting on the log, the dependence on $\mu$ is spurious. Up to coefficient  $(D-4)^2$, this operator coincides with the first one in (\ref{4d conformal}). The second operator at the second line in (\ref{4der}) at $D=4$ maps to the second operator in (\ref{4d conformal}), as expected.

Let us look now at operators with 6 derivatives in the same manner as we just did with operators with 4 derivatives. The full set of scale-invariant operators contains 7 independent ones which are not connected to each  by integrating by parts,
\begin{equation}
\label{SI}
\begin{split}
&{\cal O}_1=\frac{1}{\phi^4}\bb\phi \,\bb^2\phi,~~{\cal O}_2=\frac{1}{\phi^5}(\bb\phi)^3,~~\\ &{\cal O}_3=\frac{1}{\phi^5} \bb^2\phi\,(\dd_{\mu}\phi)^2,\\ &{\cal O}_4=\frac{1}{\phi^6} (\dd_{\nu}\phi)^2\bb(\dd_{\mu}\phi)^2,~~{\cal O}_5=\frac{1}{\phi^6} (\dd_{\mu}\phi)^2\,(\bb\phi)^2,\\ &{\cal O}_6=\frac{1}{\phi^7} (\dd_{\mu}\phi)^4\bb\phi,~~{\cal O}_7=\frac{1}{\phi^8} (\dd_{\mu}\phi)^6.
\end{split}
\end{equation}
Thus, the scale-invariant Lagrangian can be written as,
\begin{equation}
{\cal L}_6=A_1 {\cal O}_1+A_2 {\cal O}_2+\dots+A_7 {\cal O}_7. 
\end{equation}
A straightforward computation shows that the requirement of conformal invariance imposes three conditions on the coefficients $A_n$,
\begin{equation}
\begin{split}
\label{leq}
& A_5=14A_3+4 A_4-4 A_1\\
& A_6=-6(A_4+4 A_3)\\
& A_7=9 A_3-6 A_4.
\end{split}
\end{equation}
Thus, we are left with only four operators allowed by the conformal symmetry. All of them can be constructed within the method described at the beginning of this Section: just take (\ref{6der}) and put $D=6$, getting four operators
\begin{equation}
\label{4oper}
\begin{split}
{\cal O}_1=\frac{1}{\phi}\bb^3\frac{1}{\phi},~~{\cal O}_2=\frac{1}{\phi^3}\bb\phi\bb^2\log{\phi},~~\\
{\cal O}_3=\frac{1}{\phi^5}(\bb\phi)^3,~~{\cal O}_4=\frac{\bb\phi}{\phi^2}\bb\left( \frac{\bb\phi}{\phi^2} \right).
\end{split}
\end{equation}

This set is linearly connected to the set defined by conditions (\ref{leq}) as it should be. 

\subsection{Operators in $D=6$}

The number of structures appearing in the scale-invariant Lagrangian with 6 derivatives is the same in 4 and 6 dimensions (see \eqref{SI}). Thus, there are 7 scale-invariant operators and only 4 independent combinations of them are conformal. These operators can be constructed along the lines of this Section. The only subtle point in this construction is that the operator containing $\bb^3$ is vanishing in $D=6$. The way out of that is the same as in 4 dimensions and includes assuming the $1/(D-6)^2$ coefficient in front of it. In this way,
\begin{equation}
\frac{b_1}{(D-6)^2}\,\phi^{\frac{D-6}{D-2}}\square^3\left(\phi^\frac{D-6}{D-2}\right)\rightarrow \frac{b_1}{9}\log{\phi}\bb^3(\log{\phi}).
\end{equation}
\vspace{5pt}

The full set of conformal operators with 6 derivatives in $D=6$ can be written as
\begin{equation}
\begin{split}
L_6= & \frac{b_1}{9}\log{\phi}\bb^3(\log{\phi})+\\
& +b_2 \phi^{-\frac{3}{2}}\, \square^2\left(\phi^\frac{1}{2}\right)\square\phi+\\ &+b_3\phi^{-3}\, (\square \phi)^3+b_4 \phi^{-1} \square\phi\square\left(\phi^{-1}\,\square\phi\right).
    \end{split}
\end{equation}

\subsection{Operators in $D=2$}

The set of operators in \eqref{conf_O} is not well defined at $D=2$. This subsection shows how to construct all invariant operators in two dimensions.  For this end, we redefine the scalar field in such a way that there are no divergent pieces at $D=2$. Namely, if we replace $\phi$ by  $\varphi$ as follows,
\begin{equation}
    \phi=\varphi^{\frac{D-2}{2}}~,
\end{equation}
we get rid of $D-2$ in the denominator in (\ref{boxes}) and (\ref{ONhat}).  Within this definition, the field $\varphi$ has a dimension of mass to the first power in any $D$. Then, all the operators \eqref{conf_O} will have a regular limit for $D\to 2$,
\begin{equation}
  \hat{O}_N=\varphi^{-(N+1)}\square^N\varphi^{\frac{D-2N}{2}}.
\end{equation}

For $N=0$ we have a potential term in the form $\varphi^2$,  allowed by the conformal symmetry. For $N=1$ we get an operator
\begin{equation}
    \hat{O}_1=\varphi^{\frac{D-2}{2}}\square\varphi^{\frac{D-2}{2}}~,
\end{equation}
formally vanishing for $D=2$. As previously, to get a non-zero contribution we have to take
\begin{equation}
\label{kin2}
   \left( \frac{2}{D-2}\right)^2\left(\partial_{\mu}\varphi^{\frac{D-2}{2}}\right)^2=(\partial_{\mu}(\log{(\varphi/\mu)}))^2~.
\end{equation}
Here $\mu$ is arbitrary, but the action does not depend on it because of the derivative acting on the log.

At the level of 4 derivatives, we can have two operators constructed with the use of  \eqref{conf_O},
\begin{equation}
    \left(\square\left(\frac{1}{\varphi}\right)\right)^2,\quad\left(\frac{1}{\varphi}\square\log (\varphi/\mu)\right)^2~.
\end{equation}

It is convenient to redefine the scalar field further and convert the essentially non-linear kinetic term for $\varphi$ to the canonically normalised one. This can be done with the help of the transformation 
\begin{equation}
    \varphi=\mu e^{\sigma}~,
\end{equation}
where $\sigma$ is a dimensionless field. In terms of this field, the action constructed from operators (\ref{conf_O}) has the form,
\begin{equation}
\label{S_sigma}
\begin{split}
    S_{\sigma}=\int d^2 x \left(c_0 \mu^2 e^{2\sigma}+c_1(\dd\sigma)^2+\frac{c_2}{\mu^2}\left(\bb e^{-\sigma}\right)^2+\right.\\
    \left.+\frac{c_3}{\mu^2}e^{-2\sigma}(\bb\sigma)^2+\dots\right)~.
    \end{split}
\end{equation}
Since this action was found as a limit $D\to2$ of D-dimensional action, it is actually {\em not invariant} under a full conformal group which is known to have an infinite number of generators in 2 dimensions \cite{Belavin:1984vu}. The infinitesimal conformal transformation of the field $\sigma$ is a shift
\begin{equation}
    \sigma(x)\rightarrow \sigma(x')= \sigma(x) +\tau(x),\quad x'^{\mu}=x^{\mu}+\xi^{\mu}~,
\end{equation}
where $\tau$ is an arbitrary solution of the equation
\begin{equation}
\label{2dtau}
    \bb\tau=0,
\end{equation}
and $\xi^{\mu}$ is a Killing vector defined in Section 2.
The equation \eqref{2dtau} has an infinite number of solutions, $\tau=f_1(x_0+x_1)+f_2(x_0-x_1)$ with $f_{1,2}$ being arbitrary functions. 

The action \eqref{S_sigma} contains, for instance, a term with coefficient $c_2$ which is not invariant under the full conformal group. Dropping this and similar terms we arrive at the proper conformal action 
\begin{equation}
\label{dilaton_2d}
    S_{\sigma}=\int d^2 y\left[(\dd\sigma)^2+ \left(\sum_{n=0}^{\infty}c_n (\bb \sigma)^n e^{2(1-n)\sigma}\right)\right],
\end{equation}
where we introduced dimensionless coordinates $y=\mu x$. This equation can be also written in the form
\begin{equation}
\label{2d conf}
      S_{\sigma}=\int d^2 y\left[(\dd\sigma)^2+ e^{2\sigma}{\cal F}\left(e^{-2\sigma} \bb \sigma \right)\right],  
\end{equation}
where ${\cal F}$ is an arbitrary function which can be expanded in the Taylor series around the zero value of the argument. 

This form coincides with the dilaton action that can be obtained from the Weyl rescaling of the curvature invariants, see the discussion below in Sec. \ref{sec:curvinv}.

The quantum theory based on the action \eqref{2d conf} with ${\cal F}(x)\propto x^2$ was investigated in \cite{Makeenko:2021hcm}. Moreover, recent works \cite{Makeenko:2022oaf,Makeenko:2022tyu} present arguments in favor of the conjecture that the general conformal action \eqref{2d conf} actually can be reduced to the quadratic ${\cal F}(x)$, in addition to the canonical kinetic term\footnote{We thank Yuri Makeenko for discussion of these points.}.

\section{Stability of the flat direction}
The spontaneous breaking of conformal symmetry only occurs if the exact effective potential for scalar fields has a flat direction. For example, for a theory of 2 scalar fields, $\phi$ and $\chi$, the potential $V(\phi,\chi)= \chi^4  V_\chi(x)$, where $x=\phi/\chi$ has to obey the following properties \cite{Shaposhnikov:2008xi}: $V_\chi(x_0)=V'_\chi(x_0)$ with finite $x_0$. If true, the theory has a degenerate ground state and the dilaton is a Goldstone boson of the broken dilatation symmetry having zero mass. 

At the level of effective action constructed for the dilaton in Section \ref{sec:effact} the flat direction is present if the coefficient $\lambda$,  in front of the operator without derivatives, $\phi^4$, vanishes. 
Within the procedure described in Section \ref{sec:effCFT}, the choice $\lambda=0$ is perturbatively stable for the dilaton effective action. 

Indeed, the full dilaton action in D dimensions has the form
\begin{equation}
\label{fulact}
L=\frac{1}{2}(\dd\phi)^2-\frac{\lambda}{4}\phi^{2 D/(D-2)} + ...
\end{equation}
where dots represents all operators (\ref{conf_O}) with more than two derivatives. In non-zero dilaton background, the dilaton mass $m_D$ is proportional to $\lambda$ and is given by
\begin{equation}
\label{dmass}
m_D^2 = \lambda\frac{D(D+2)}{2(D-2)^2}\phi^{4/(D-2)}~.
\end{equation}
This means that all the perturbative loop corrections to the effective potential in DimReg necessarily contain powers of $\lambda$ and are zero when $\lambda=0$.

It is difficult to say what happens at the non-perturbative level. We present below a consideration based on a specific resummation inspired by the one-loop Coleman-Weinberg construction of the effective potential \cite{Coleman:1973jx}. 

The standard expression for the one-loop effective action reads,
\begin{equation}
S_1=-\frac{i}{2} \log{{\rm det}\left(\frac{\delta^2 L}{\delta\phi^2}\right)}\,,
\end{equation}
leading to an effective potential for background field $\phi$ in our theory (\ref{fulact}) with $\lambda=0$:
\begin{equation}
\label{effpot}
\begin{split}
V=\frac{1}{2} \int \frac{d^D k}{(2\pi)^D}\log\left[k^2(1+F(k^2/\phi^{4/(D-2)}))\right]=\\
\frac{1}{2} \int \frac{d^D k}{(2\pi)^D}\log\left[1+F(k^2/\phi^{4/(D-2)})\right]~.
\end{split}
\end{equation}
Here the function $F$ is determined by the higher dimensional operators (\ref{conf_O}).
We used DimReg and Euclidean space-time. 

On dimensional grounds 
\begin{equation}
\label{comp}
V= \kappa[F,D] \phi^{2D/(D-2)}~,
\end{equation}
where $ \kappa$ is controlled by the dimension of space-time and function $F$. Depending on the choice of the coefficients in front of higher-dimensional operators leading to a specific function $F$, different scenarios are possible. 

\begin{itemize}
\item If the integral (\ref{effpot}) is convergent for $D=4$ (take, for instance, $F(x) =\exp(-x^2)$), then  $\kappa[F,D]$ is finite, meaning that the $\phi^4$ interaction will be generated with computable coefficient\footnote{This type of function may appear in the ghost-free non-local theories \cite{Koshelev:2020fok,Koshelev:2021orf}.}. 
\item If $\kappa[F,D]$ contains poles in $D-4$, the scalar self-interaction is generated but it can be consistently set to zero using the renormalization procedure. This case would happen, for instance, if only one higher-dimensional operator,  such as 
\begin{equation}
a \frac{(\bb \phi)^2}{\phi^{\frac{4}{D-2}}}.
\end{equation}
is added to the kinetic term.
\item If $\kappa[F,D]=0$, the self-interaction is not generated by the resummed one-loop quantum corrections. This option can be realised if $1+F(x) = \exp(P(x))$, where $P(x)$ is an arbitrary finite polynomial.
\end{itemize}

\section{Conformal operators from curvature invariants}
\label{sec:curvinv}
Up to now all of our considerations were carried out in the flat space-time.  Since conformal symmetry acts as a Weyl transformation of metric with a specific rescaling factor (see eq. (\ref{flatct}), it is closely related to the Weyl symmetry.  In D-dimensional curved space-time, the Weyl transformations of the metric and a scalar field are defined as
\begin{equation}
\label{weylt}
g_{\mu\nu}\rightarrow \Omega^2 g_{\mu\nu},~\phi\to\Omega^{\Delta}\phi~,
\end{equation}
where $\Omega$ is an arbitrary function of space-time. The theory is Weyl invariant if the transformation (\ref{weylt}) does not change its action.

This connection between the conformal and Weyl symmetries provides yet another way to construct conformally invariant operators in flat space-time (see, e.g. \cite{Komargodski:2011vj,Luty:2012ww,Baume:2013ika,Osborn:2015rna}).  One can take an arbitrary Diff-invariant action constructed from the metric only (i.e. consider pure gravity) and replace the metric $g_{\mu\nu}$ by $g_{\mu\nu}\phi^{2/\Delta}/M_P^2$, where $M_P$ is any scale normally taken to coincide with the Planck mass. One gets in this way a scalar-tensor gravity which is Weyl-invariant by construction. Now, it is obvious that a Weyl-invariant theory is automatically conformal invariant if the metric is taken to be flat (for original discussion see  \cite{Zumino:1970tu}). 

It is instructive to see that this procedure indeed allows finding the conformally invariant operators obtained previously. For several operators, related to the Weyl anomaly, it is important to start from fractional $D$ and eventually take a limit $D\to 4$ resolving possible singularities. We will carry out this consideration for a few operators in $D=4$ and the complete dilaton action in $D=2$.

\subsection{D=4}
    
Let us start with the curvature operator having 2 derivatives of the metric,
\begin{equation}
    L_2=A\, R.
\end{equation}
Applying the described procedure to this Lagrangian and omitting total derivatives, one obtains the conformally invariant dilaton kinetic term, 
\begin{equation}
    L{\phi}=-6 A\phi\square \phi=6 A (\partial_{\mu}\phi)^2~.
    \end{equation}
This matches the previous consideration of Section \ref{sec:effact}.

The operators with 4 derivatives are expected to appear from the quadratic curvature invariants.
There are three of them,
\begin{equation}
    L_4=C_1 R^2+C_2 W_{\mu\nu\lambda\rho}W^{\mu\nu\lambda\rho}+C_3 E_4~.
\end{equation}
Here
\begin{equation}
    E_4=W_{\mu\nu\lambda\rho}W^{\mu\nu\lambda\rho}+\frac{2}{3}R^2-2 R_{\mu\nu}R^{\mu\nu}
\end{equation}
is the Euler density operator and $W_{\mu\nu\lambda\rho}$ is the Weyl tensor. As we know, there are two independent conformal operators with 4 derivatives \eqref{4d conformal}. And, indeed, the term $W^2$ cannot produce any non-trivial dilaton operators, so the only relevant terms are $R^2$ and $E_4$. The first of them  leads to the dilaton action
\begin{equation}
36 C_1 \frac{(\square \phi)^2}{\phi^2},
\end{equation}
matching the second operator in (\ref{4d conformal}). The first operator in (\ref{4d conformal}) can be obtained as a formal limit
\begin{equation}
\lim_{D\to 4} E_4/(D-4)^2\rightarrow 4 \log{\phi}\bb^2\log{\phi}~.
\end{equation}
Note that this expression was obtained up to total derivatives (in arbitrary $D$). The term $E_4/(D-4)$ would also give a total derivative lagrangian for the scalar field. 

The reason for this singular prescription is that $E_4$ is a total derivative in $D=4$. If we were to construct conformally invariant operators with this method directly in $D=4$, we would miss this operator. This mismatch in the number of operators is connected to the Weyl anomaly. 

For more than 4 derivatives, the number of different curvature invariants and their contractions grows and becomes larger than the number of conformal operators. We expect that all conformal operators with more than 4 derivatives can be obtained from the curvature invariants without any singular terms in 4 dimensions. It is important to stress that this statement does not contradict the well-known mathematical fact that there are no higher derivative Weyl invariant kinetic terms for a scalar field with positive Weyl weight \cite{FG,GG,G}.  In the field theory language, the results of \cite{FG,GG,G}  tell that in the space of even $2k$ integer dimensions the {\em quadratic in field $\Phi$} action in the form $\Phi \Box^l \Phi$ can be extended in Weyl-invariant way to generic curved space-time only if $l \leq k$.  The procedure outlined at the beginning of this Section produces the actions which are {\em not quadratic} in the scalar field, explaining why the theorem of \cite{FG,GG,G}  is not applicable.

We illustrate this conjecture for operators with 6 derivatives. There are 5 curvature invariants with 6 derivatives which do not contain the Weyl tensor (those with the Weyl tensor are irrelevant as $W$ is zero on the conformally flat metric $\eta_{\mu\nu}\phi^{2/\Delta}/M_P^2$). They are collected in the action
\begin{equation}
\label{box3}
\begin{split}
    L_6=A_1 R^3+A_2 R_{\mu\nu}^2R+A_3 R_{\mu\nu}R^{\nu}_{\lambda}R^{\lambda\mu}+\\
    +A_4 \nabla_{\mu}R\nabla_{\mu}R+A_5\nabla_{\mu}R_{\nu\rho}\nabla_{\mu}R^{\nu\rho}.
\end{split}
\end{equation}
The straightforward but tedious computation shows that this action produces four independent conformal operators (as it should be, see Section \ref{sec:effact}),
\begin{equation}
\label{Bdef}
    L_c=B_1 {\cal O}_1+B_2{\cal O}_2+B_3{\cal O}_3+B_4{\cal O}_4,
\end{equation}
where the operators ${\cal O}_i$ are defined in (\ref{4oper}).

There is no contradiction in the fact that 5 curvature operators are mapped to 4 conformal scalar field operators because one specific combination, namely, 
\begin{equation}
\begin{split}
    L_b\propto  R^3-7 R_{\mu\nu}^2R+12 R_{\mu\nu}R^{\nu}_{\lambda}R^{\lambda\mu}-\\
    -2 \nabla_{\mu}R\nabla_{\mu}R+6\nabla_{\mu}R_{\nu\rho}\nabla_{\mu}R^{\nu\rho},
    \end{split}
\end{equation}
is a  total derivative on the conformally flat metric. The values of $A_i$ and $B_i$ are connected as
\begin{equation}
\begin{split}
    B_1 = \frac{A_3}{2} - A_5,\\B_2 = 2 (6 A_2 + 5 A_3 - 3 A_5),\\ B_3 = 
 -2 (108 A_1 +30 A_2 + 9 A_3 - 18 A_4 - 7 A_5),\\ B_4 = 
 -\frac{1}{2} (24 A_2 + 21 A_3 + 72 A_4 + 10 A_5).
 \end{split}
\end{equation}
For completeness, we also give the inverse relations: 
\begin{equation}
\label{Adef}
\begin{split}
A_2 = -7 A_1 + \frac{1}{216} (81 B_1 - 24 B_2 - 7 B_3 - 7 B_4),\\ 
A_3 = 12 A_1 - \frac{1}{18} (27 B_1 - 6 B_2 - B_3 - B_4), \\
A_4 = -2 A_1 + \frac{1}{108} (60 B_1 - 9 B_2 - B_3 - 4 B_4), \\
A_5 =  6 A_1 - \frac{1}{36} (63 B_1 - 6 B_2 - B_3 - B_4).
 \end{split}
\end{equation}
In this way, we can present one-to-one explicit correspondence between the curvature invariants and conformal operators.

 The Weyl invariant action involving the scalar field $\phi$ in a non-polynomial way and the arbitrary metric $g_{\mu\nu} $ can be constructed from eq. (\ref{box3})  by replacing $g_{\mu\nu} \to g_{\mu\nu}\phi^{2/\Delta}/M_P^2$  and using the rules for conformal transformations of the curvature tensors given, for instance, in \cite{Nakayama:2013is}. The explicit formula for the 
 Weyl-invariant generalisation of the most interesting part of the action associated with the operator  ${\cal O}_1$ defined in eq. (\ref{4oper}) is given in Appendix.

\subsection{D=2}
In $D=2$ the Riemann and Ricci tensors can be expressed through the Ricci scalar. For this reason, the relevant Lagrangian that can be used to construct the dilaton action can be written as
\begin{equation}
    L=\sum_{n=0}^{\infty}c_n R^n~.
\end{equation}
Substituting $g_{\mu\nu}=e^{2\sigma}\eta_{\mu\nu}$ on can recover the general form of the dilaton action given in \eqref{dilaton_2d}. Note that the first term here is a result of the limit $\lim_{D\to 2} \frac{1}{D-2} R$, associated with the fact that in $D=2$ the scalar curvature $R$ is a total derivative, in analogy with $E_4$ invariant in four dimensions.

\section{\label{sec:2} Weyl anomaly in curved space-time and conformal symmetry}
\subsection{Weyl anomaly}
\label{sec:anomaly}
The discussion of the previous section shows that for a fractional number of dimensions $D$ all conformally invariant operators in flat space-time can be constructed from Weyl-invariant action taking its flat space limit. If the number of dimensions is odd, $D=2k+1$ and $k$ is an integer, the limit $D \to 2k+1$ is regular, and any conformal operator allows a Weyl-invariant extension. If the number of dimensions is even,  $D=2k$, all conformally invariant operators with $N\neq 2k$ derivatives also allow a regular Weyl-invariant extension, meaning that only one operator, with $N= 2k$ derivatives, presents an obstruction for the building of the Weyl-invariant action.  For example, the Weyl-invariant curved space extension of the flat space conformally invariant action 
\begin{equation}
\label{box2}
    S=\int{d^D x  \phi^{\frac{4-D}{D-2}}\Box^2\left(\phi^{\frac{4-D}{D-2}}\right)},
\end{equation}
for $D\to 4$ is given by
\begin{widetext}
\begin{equation}
\label{div}
\int d^4 x \left(\tau\bb^2\tau\right) \to S=\lim_{D\to 4} \int d^D x \sqrt{-g}\left[\tau\Delta_4\tau+
2\tau \left(-\frac{1}{6} \bb R +\frac{1}{4} E_4\right)+\frac{R^2}{36} + L_{anom}\right]~, 
\end{equation}
\end{widetext}
where $\tau=\log(\phi/\mu)$, $\Delta_4$ is the Riegert operator \cite{Fradkin:1981jc, paneitz, Riegert:1984kt},
\begin{equation}
 \Delta_4=\bb^2+2R^{\mu\nu}\nabla_\mu\nabla_\nu-\frac{2}{3}R\bb+\frac{1}{3}(\nabla^\mu R) \nabla_\mu
\end{equation}
and
\begin{equation}
\label{anom}
 L_{anom}=\frac{E_4}{2(D-4)}~.
\end{equation}
The Euler density $E_4$ is a total derivative only in $D=4$ so it cannot be dropped if $D\neq 4$. This is exactly the manifestation of the Weyl anomaly \cite{Duff:1993wm}.

Given the fact that the term $\tau \Box^2\tau$ will inevitably appear in an effective theory for the dilaton \cite{Komargodski:2011vj},  we cannot avoid the singularity which is required to make the action Weyl-invariant.  In other words, Weyl symmetry cannot survive if quantum effects leading to the generation of the arbitrary dilaton effective action are accounted for.  (However, {\em non-local} operators providing the Weyl invariant action can be constructed \cite{Fradkin:1983tg, Antoniadis:1991fa}.) This makes it clear that the extension of the flat space conformal invariance to the curved space cannot be the Weyl symmetry if the dimension of space-time is four and the locality of the action is imposed.

A similar situation shows up in all even dimensions. For example, the Weyl invariant extension of the operator $\sigma\bb\sigma$ requires adding the term $R/(D-2)$ in 2 dimensions which is singular when $D\to 2$. In the same way, in 6 dimensions the Weyl-invariant extension of the operator $\log{\phi}\bb^3(\log{\phi})$ will require the term $E_6/(D-6)$ where $E_6$ is the 6-dimensional Euler density which is 
cubic in curvature tensor and is total derivative in 6 dimensions, see \cite{Osborn:2015rna} for the explicit expression.

Though the Weyl symmetry cannot be realised in quantum theory, it is natural to ask whether one can find some subgroup of the Weyl group which is anomaly-free and matches with the flat-space conformal symmetry. We will show that the only anomaly-free subgroup is that of global scale transformations, meaning that the full conformal group in the curved space-time is broken down to this symmetry due to quantum effects.

\subsection{Anomaly-free condition in $D=4$}

As we see, the absence of the regular Weyl-invariant extension of the term \eqref{box2} leads to the conclusion that Weyl symmetry cannot be made anomaly-free in four dimensions. Can the smaller subgroup of Weyl transformations be kept free from the quantum anomaly? In this Section, we derive the condition on $\omega$ for the infinitesimal Weyl transformation $\Omega=1+\omega$.

A small Weyl transformation ($1+\omega$) acts on the term \eqref{box2} (which is the only source of the Weyl anomaly in $D=4$) in the limit $D\rightarrow 4$ as,
\begin{equation}
\label{deltabox2}
\delta_{\omega} S=\int{d^4 x\sqrt{-g}\left(\nabla_{\alpha}\omega \nabla^{\alpha}\tau\Box\tau+\frac{1}{2}\Box{\omega}\Box{\tau}\right)}.
\end{equation}
Recall that here $\tau=\log{\phi/\mu}$.
The first term can be written in a bit different form by means of omitting the total derivative,
\begin{equation}
\label{deltabox2dd}
    \delta_{\omega} S=\int{d^4 x\sqrt{-g}\left(-\nabla_{\alpha}\nabla_{\beta}\,\omega \nabla^{\alpha}\tau\nabla^{\beta}\tau+\frac{1}{2}\Box{\omega}\Box{(\tau^2)}\right)}.
\end{equation}
One can see that the condition
\begin{equation}
\label{anomaly-free}
    \nabla_{\mu}\nabla_{\nu}\omega=0
\end{equation}
is  {\it sufficient} to make the symmertry restricted by the condition \eqref{anomaly-free} anomaly-free.

What does this condition mean for the curved space with an arbitrary metric? Differentiating the condition \eqref{anomaly-free} and taking the commutator, one can find,
\begin{equation}
[\nabla_{\rho}\nabla_{\mu}]\nabla_{\nu}\omega =R_{\mu\rho\sigma\nu} \nabla^{\sigma} \omega=0.
\end{equation}
All solutions of the equation \eqref{anomaly-free} must satisfy $R_{\mu\nu}\nabla^{\nu}\omega=0$. On top of an arbitrary spacetime, this equation has only a trivial solution $\nabla_{\sigma}\omega=0$ which means that $\omega=const$ reflecting the fact that only dilatation symmetry can be kept anomaly-free. 

 The condition \eqref{anomaly-free} guarantees the absence of the anomaly but it can be too strong. Can it be weaker in fact, at least for some special background metrics? The problematic term $\tau \square^2 \tau$ can be written in a Weyl-invariant form \eqref{div}.
Recall that the only problem of preserving the Weyl symmetry anomaly-free is related to the term $L_{anom}$ which will bring extra divergence when making the counter term for $\tau\bb^2\tau$ Weyl-invariant. If our requirement is weaker and we want to preserve only some class ${\cal M}_s$ of Weyl transformations free from anomaly then we have to require that for $\omega \in {\cal M}_s$ the transformation of $L_{anom}$,
\begin{equation}
\label{varanom}
 \delta_{\omega} L_{anom}=\frac{1}{2}E_4 \omega 
\end{equation}
leads to a surface integral in the action,
\begin{equation}
\label{intan}
\int d^4 x \sqrt{-g} \frac{1}{2}E_4\omega.
\end{equation}
This can happen only if the variation of (\ref{intan}) with respect to the metric  $g_{\mu\nu}\to g_{\mu\nu} +h_{\mu\nu}$ is zero up to a total derivative.

A straightforward computation of the perturbed action \eqref{intan} up to the total derivative gives
\begin{equation}
    \delta E_\omega = \int d^4 x \sqrt{-g}h_{\mu\nu}\Sigma^{\mu\nu\alpha\beta}\nabla_{\alpha}\nabla_{\beta}\,\omega~,
\end{equation}
where
\begin{widetext}
\begin{equation}
\Sigma^{\mu\nu\alpha\beta}=2R(g^{\alpha\mu}g^{\beta\sigma}-g^{\alpha\beta}g^{\mu\nu})+4 R^{\mu\nu}g^{\alpha\beta}+4g^{\mu\nu}R^{\alpha\beta}-8 g^{\mu\beta}R^{\alpha\nu}-4 R^{\mu\alpha\nu\beta}~.
\end{equation}
\end{widetext}
Thus, the {\it nessesary} condition on $\omega$ which can make the symmetry anomaly-free reads,
\begin{equation}
\label{sigma}
\Sigma^{\mu\nu\alpha\beta}\nabla_{\alpha}\nabla_{\beta}\omega=0~.
\end{equation}

Recall that we consider theories with dynamical gravity which means that we are dealing with a dynamical metric of spacetime. If the metric is arbitrary, the condition (\ref{sigma}) can be satisfied only for $\nabla_{\mu}\nabla_{\nu}\omega=0$ which brings us back to the condition \eqref{anomaly-free}.


One may wonder if there can exist some peculiar geometries allowing for solutions of  \eqref{sigma} with $\omega \neq const$. We were not able to find any. For example, for the maximally symmetric space with $R^{\mu\alpha\nu\beta}=R_0(g^{\mu\nu}g^{\alpha\beta}-g^{\mu\beta}g^{\nu\alpha})$ the equation \eqref{sigma} turns to be
\begin{equation}
    -4R_0(g^{\mu\nu}\bb\omega-\nabla^{\mu}\nabla^{\nu}\omega)=0,
\end{equation}
which is equivalent to $\nabla^{\mu}\nabla^{\nu}\omega=0$, leading again to the solution $\omega=const$. 

Let us stress here that the equation \eqref{sigma} is the necessary condition which is not sufficient to claim the absence of the conformal anomaly on top of some specific geometry. However, as we have shown, even this condition cannot be satisfied either for the general metric in dynamical gravity or the maximally symmetric spaces. The only exception corresponds to the case of the flat spacetime for which we obtain that conformal symmetry can be anomaly-free which is fully consistent with the results of Section \ref{sec:effCFT} obtained in a way which does not involve gravity and curved space. 

A remark is now in order. Our result says that there are no anomaly-free subgroups of Weyl transformations with dynamical gravity, except dilatations. There are two larger subgroups of the Weyl symmetry considered in the literature \cite{Iorio:1996ad,Karananas:2015ioa,Edery:2014nha,Edery:2015wha}. Covariant extension of the conformal transformations in curved space can be defined \cite{Iorio:1996ad,Karananas:2015ioa} (see also \cite{Rychkov:2016iqz}) via the Killing vector $\xi_{\mu}$ satisfying the equation
\begin{equation}
\label{killing1}
 \nabla_{\mu}\xi_{\nu}+\nabla_{\nu}\xi_{\mu}=\frac{2}{D}g_{\mu\nu}\nabla^{\alpha}\xi_{\alpha}~.
\end{equation}
The corresponding infinitesimal Weyl factor is given by $\omega=\nabla_{\alpha}\xi^{\alpha}$.  For the transformations \eqref{killing1} 
\vspace{5pt}
\begin{equation}
\begin{split}
\nabla_{\alpha}\nabla_{\beta}\,\omega=\frac{D}{2-D}\left(\xi^{\mu}\nabla_{\mu}R_{\alpha\beta}+\frac{2}{D}\omega R_{\alpha\beta}+\right.
    \\
    \left.+\frac{g_{\alpha\beta}}{2(1-D)}\left(\frac{2}{D}\omega R+\xi^{\mu}\nabla_{\mu}R\right) \right),
    \end{split}
\end{equation}
which is, in general, non-zero for the arbitrary choice of the metric, meaning that this symmetry is anomalous.
Another related possibility extending the conformal symmetry was named restricted Weyl transformations \cite{Edery:2014nha,Edery:2015wha}. In these works, it was shown that in $D=4$ all the transformations satisfying $\Box \Omega=0$ form a subgroup in the full group of all Weyl transformations, with respect to the operation of their composition. In an arbitrary number of dimensions the similar condition,
\begin{equation}
\label{restricted}
    \Box \left(\Omega^{\frac{D-2}{2}}\right)=0,
\end{equation}
highlights the same subgroup structure. Still, this symmetry cannot be preserved at the quantum level because this condition is weaker than $\nabla_{\mu}\nabla_{\nu} \Omega=0$ which would grant the absence of the anomaly. 

\section{Scale-invariant Lagrangian for the dilaton and gravity}

In this Section, we present a Lagrangian for the dilaton and gravity in $D=4$ which is invariant under the anomaly-free scale transformations. All terms that could be written with two derivatives are
\begin{equation}
 \label{SWeyl-inv2}
 S=\int{d^4 x\sqrt{-g}\left[-\frac{1}{2}\zeta\phi^2 R+\frac{1}{2}(\partial_{\mu}\phi)^2-\frac{\lambda}{4}\phi^4\right]},
\end{equation}
where $\zeta$ and $\lambda$ are arbitrary constants (note that Weyl invariance would impose a specific value for the non-minimal coupling $\zeta=-1/6$). Making a transition to the Einstein frame one can see that the combination $\frac{\lambda M_P^4}{\zeta^2}$ is the vacuum energy. In these terms, the cosmological constant problem is converted to the question of why $\lambda/\zeta^2\ll 1$.

 The general Lagrangian invariant under the scale transformations and respecting parity at the level of four derivatives can be written as
 \vspace{10pt}
\begin{widetext}
\begin{equation}
\label{SWeyl-inv4}
\begin{split}
 S=\int d^4 x \sqrt{-g} \left[A R\Box \tau+B R(\partial_{\mu}\tau)^2)+C G^{\mu\nu}\partial_{\mu}\tau\partial_{\nu}\tau + F\tau E_4 +E R^2+E W_{\mu\nu\lambda\rho}^2+G((\partial_{\mu}\tau)^2)^2\right.+\\
 \left.+H(\Box \tau)^2+J (\Box \tau+(\partial_{\mu}\tau)^2)^2\right]~ .
 \end{split}
\end{equation}
\end{widetext}
Here $A,B,C,E,F,G,H$ and $J$ are arbitrary constants, $G_{\mu\nu}$ is the Einstein tensor, $\tau=\log({\phi/\mu})$ with $\mu$ being an arbitrary scale which is not relevant in perturbation theory. The field $\tau$ transforms under the dilatations as $\tau\rightarrow \tau-\omega$. 

The second line of \eqref{SWeyl-inv4} contains only the operators which are allowed by the conformal symmetry in the flat space limit. Given the fact that the conformal symmetry is broken by gravity, the operators in the first line are expected to be suppressed by the Planck scale. This could not be the case for the two conformal operators since they have an enhanced symmetry in the flat space limit.

The structure of the action given by (\ref{SWeyl-inv2},\ref{SWeyl-inv4}) allows also us to clarify the situation with the energy-momentum tensor. It is well-known \cite{Callan:1970ze} that in CFT in flat space it is possible to define an (improved) stress-energy tensor $T_{\mu\nu}$ with zero trace, $T_\mu^\mu=0$. If the theory were Weyl invariant, this relation would remain in force in the gravitational background. However, when the quantum corrections are incorporated in a Weyl-invariant way in $D$-dimensions as in \cite{Englert:1976ep,Shaposhnikov:2008xi}, and the limit $D\to 4$ is taken, $T_\mu^\mu$ receives several contributions containing Diff invariants $W^2,~E_4,~R^2$ and $\bb R$ \cite{Duff:1993wm}. One of them ($E_4$, the so-called a-anomaly) cannot be removed by the Weyl-invariant counter-term, signalling that the Weyl symmetry is anomalous \footnote{Note that the term $W^2$ representing the so-called c-anomaly can be taken away by the Weyl-invariant counter-term  $\phi^{\epsilon} W^2/ \epsilon$ \cite{Englert:1976ep}.}. The scale symmetry in curved space {\em does not impose} that the trace of stress-energy tensor is zero, only a weaker condition $\int d^4x \sqrt{-g} T_\mu^\mu \omega=0$, where $\omega=$const  must be satisfied.

The action \eqref{SWeyl-inv4} contains only parity-even operators. If we relax this assumption, the dilatation symmetry will allow us to write more operators. In particular, among terms containing only dilaton and gravity with 4 derivatives, we can have the coupling between the dilaton and Pontryagin density,
\begin{equation}
    L_{CP}= \kappa \tau \epsilon_{\alpha\beta\gamma\delta}R^{\alpha\beta\rho\sigma}R_{\rho\sigma}^{~~\gamma\delta},
\end{equation}
known also as the gravitational Chern-Simons term ($\kappa$ is an arbitrary dimensionless constant). 

Another phenomenologically interesting possibility is connected to the fact that the dilaton can be also coupled to the non-abelian gauge fields via the Chern-Simons term,
\begin{equation}
\label{axion}
L_{gauge}=\gamma \tau {\rm Tr}(G_{\mu\nu}\tilde{G}^{\mu\nu}).    
\end{equation}
Here $\gamma$ is an arbitrary dimensionless constant. In this case, if the gauge theory is in the strong coupling regime, the scale symmetry will be broken by non-perturbative effects, similarly to the QCD axion models, see \cite{GrillidiCortona:2015jxo} for a review. Thus, dilaton becomes massive. Moreover, the dilaton coupled to the SM gauge fields can be relevant for solving the strong CP problem acting in a way similar to the axion \cite{Peccei:1977hh,Kim:1979if,Shifman:1979if,Dine:1981rt}. If the dilaton field $\tau$ is defined in such a way that it transforms to $-\tau$ under the parity transformation (i. e. it is a pseudoscalar) then the coupling \eqref{axion} resembles the effective action of the QCD axion while preserving the CP symmetry\footnote{Note that, within this choice of dilaton transformation, some terms in \eqref{SWeyl-inv4} are forbidden, $A=F=J=0$ if one requires CP invariance}. The pseudoscalar dilaton would correspond to some generalization of the original scale symmetry supplemented by the parity transformation. The phenomenology of this dilaton model is discussed in a cosmological context in the recent work \cite{Belokon:2022pqf}.

\section{Conclusions and discussion}

In this paper, we address the question of whether conformal symmetry can be a fundamental symmetry of Nature which is spontaneously broken at low energies. To this end, we first considered a theory in flat space-time and elucidated the effective theory construction allowing conformal symmetry to be intact in all orders of perturbation theory. We constructed the most general effective action for the dilaton in a flat space of arbitrary dimension and clarified the connection of our method with that which uses the curved space and curvature invariants. The spontaneous breaking of conformal invariance requires the existence of the flat direction in the effective potential for the dilaton. We discussed the stability of this direction with respect to perturbative quantum corrections associated with the dilaton field itself. 

Conformal symmetry was originally defined and widely used for field theories in flat space without gravity. However, since any realistic theory should contain gravity we need to understand the extension of this symmetry for dynamical gravity and curved spacetime metric. A natural candidate for this extension is a Weyl group or its (finite) subgroups. We have shown that the only anomaly-free subgroup of the Weyl transformations is that of the global scale symmetry, meaning that the conformal symmetry is necessarily broken down to dilatations by the gravitational effects. We constructed the dilaton-gravity action up to the terms with four derivatives respecting parity. 

Talking about phenomenological and cosmological applications, the graviton-dilaton action we presented can be complemented by all the fields of the Standard Model or $\nu$MSM \cite{Asaka:2005an,Asaka:2005pn} in a scale-invariant invariant way, the explicit equations can be found in \cite{Shaposhnikov:2008xb}. In this work, we justify that only the scale invariance is a symmetry which can be preserved at the quantum level in realistic theories containing gravity. Our findings reveal that scale invariance implies conformal invariance at the level of the lowest order action in the flat space, however, when gravity is included, the symmetry of the Higgs-dilaton action reduces back only to the subgroup of dilatations. However, the higher-order operators which had conformal symmetry in the flat space, are expected to be less suppressed than those which have only scale symmetry. This hierarchy is a key observation that is relevant for phenomenology since the cutoff scale in the scalar sector in the Higgs-dilaton model is known to be much less than the Planck scale.

It has been shown that this Higgs-dilaton Lagrangian can solve all the observational problems of the Standard Model (such as inflation, neutrino masses, baryon asymmetry of the Universe, and Dark Matter), provided the scale symmetry is spontaneously broken (for discussion of inflation see \cite{Garcia-Bellido:2011kqb}, and for review of other problems \cite{Boyarsky:2009ix}). It is of crucial importance that the non-minimal couplings of the Higgs and dilaton fields to the Ricci scalar are not constrained by any value ($-\frac{1}{6}$ for the Weyl symmetry). This freedom allows to arrange the appearance of all scales in the low energy physics and gravity (namely, Higgs vev and Planck mass) from one source -- dilaton vev. We also briefly discuss the possibility to solve the strong CP problem in QCD using the dilaton Chern-Simons coupling to the gauge fields and non-perturbative breaking of the scale symmetry. We underline once more that the massless dilaton does not lead to the fifth force and thus is harmless from the experimental point of view \cite{Wetterich:1987fm, Wetterich:1987fk,Shaposhnikov:2008xb,Ferreira:2016kxi}.

We thank Georgios Karananas, Andrei Mikhailov,  Alexander Monin, Emil Mottola, Valery Rubakov and Arkady Tseytlin for useful discussions. A.T. is supported by Simons Foundation Award ID 555326 under the Simons Foundation Origins of the Universe initiative, Cosmology Beyond Einstein’s Theory. A.T. thanks EPFL, where a part of this work was done, for hospitality. This work was supported by ERC-AdG-2015 grant 694896 and by the Swiss National Science Foundation Excellence grant 200020B 182864.  M.S. was also supported  by the Generalitat Valenciana grant PROMETEO/2021/083.

\section{Appendix}
In this Appendix, we construct explicitly a Weyl-invariant extension of the action 
\begin{equation}
S=\int d^4 x {\cal O}_1=\int d^4 x \frac{1}{\phi}\bb^3\frac{1}{\phi}~.
\end{equation}
to an arbitrary space-time. To simplify the formulas, we introduce $\Phi=\frac{1}{\phi}$ with the canonical mass dimension GeV$^{-1}$. 

It is known from \cite{FG,GG,G} that the Weyl-covariant generalisation of the linear operator $\bb^3$ does not exist in 4-dimensional space-time if the Bach tensor is non-zero (see Appendix A of  \cite{Osborn:2015rna} for definition and helpful discussions).  The procedure formulated in Section \ref{sec:curvinv} leads to essentially non-quadratic action, meaning that there is no contradiction with the non-existence theorem. 

The operator  ${\cal O}_1$ is singled out by the choice $B_1=1,~B_2=B_3=B_4=0$ in eq. (\ref{Bdef}). As follows from eq. (\ref{Adef}),  its Weyl-invariant generalisation is derived from the following combination of the curvature invariants:
\begin{equation}
\label{actO1}
S_{\cal O}=\frac{7}{24} R^3-\frac{5}{3} R R_{\mu\nu}R^{\mu\nu}+2R_{\mu\nu}R^{\nu\rho}R_\rho^\mu-\frac{1}{36}R_{;\mu}R^{;\mu}~.
\end{equation}
Replacing the metric  $g_{\mu\nu}$ in (\ref{actO1}) by $g_{\mu\nu}\phi^{2/\Delta}/M_P^2$ one finds
\begin{equation}
\label{actO2}
S_{\cal O}= S_1+S_2+S_3~,
\end{equation}
where
\[
S_1=\Phi^{\,\,;\mu}_{;\mu\,\,;\nu}\Phi_{;\rho}^{\,\,;\rho;\nu}
+\frac{16}{\Phi} \Phi_{;\mu;\nu}\Phi^{;\mu;\nu}\Phi_{;\rho}^{;\rho}
\]
\[
-\frac{3}{\Phi}\Phi_{;\mu}^{;\mu}\Phi_{;\nu}^{;\nu}\Phi_{;\rho}^{;\rho}
-\frac{16}{\Phi} \Phi_{;\mu;\nu}\Phi^{;\mu}_{;\rho}\Phi^{;\nu;\rho}
\]
\[
+\frac{2}{\Phi}\Phi_{;\mu}^{;\mu}\Phi^{;\nu}\Phi^{\,\,;\rho}_{;\rho\,\,;\nu}
-\frac{8}{\Phi}\Phi^{;\mu}\Phi_{;\mu}^{;\nu}\Phi^{\,\,;\rho}_{;\rho\,\,;\nu}
\]
\[
-\frac{8}{\Phi^2}\Phi_{;\mu}\Phi^{;\mu}\Phi_{;\nu;\rho}\Phi^{;\nu;\rho}
+\frac{16}{\Phi^2}\Phi^{;\mu}\Phi^{;\nu}\Phi_{;\mu;\rho}\Phi_{;\nu}^{;\rho}
\]
\[
-\frac{8}{\Phi^2}\Phi^{;\mu}\Phi^{;\nu}\Phi_{;\mu;\nu}\Phi_{;\rho}^{;\rho}
+\frac{3}{\Phi^2}\Phi_{;\mu}\Phi^{;\mu}\Phi_{;\nu}^{;\nu}\Phi_{;\rho}^{;\rho}~,
\]
\[
S_2=-\frac{7}{24} R^3\Phi^2 
+\frac{1}{36}R_{;\mu}R^{;\mu}\Phi^2
\]
\[
+\frac{1}{9}R R_{;\mu}\Phi \Phi^{;\mu}
+\frac{11}{18} R^2 \Phi_{;\mu} \Phi^{;\mu}
-\frac{23}{12}R^2\Phi\Phi_{;\nu}^{;\nu}
\]
\[
-\frac{4}{3} R_{;\mu}\Phi^{;\nu}\Phi_{;\nu}^{;\mu}
+\frac{1}{3}R_{;\mu}\Phi^{;\mu}\Phi_{;\nu}^{;\nu}
-\frac{25}{6}R\Phi_{;\mu}^{;\mu}\Phi_{;\nu}^{;\nu}+
\]
\[
+\frac{20}{3}R \Phi_{;\mu;\nu}\Phi^{;\mu;\nu}
+\frac{1}{3}R_{;\mu}\Phi\Phi_{;\nu}^{\,\,;\nu;\mu}
\]
\[
+\frac{2}{3}R \Phi_{;\mu}\Phi_{;\nu}^{\,\,;\nu;\mu}
+\frac{20}{3}R R_{\mu\nu}\Phi\Phi^{;\mu;\nu}
\]
\[
+16 R_{\mu\nu}\Phi^{;\mu;\nu}\Phi_{;\rho}^{;\rho}
-12 R_{\mu\nu}R^\nu_\rho\Phi\Phi^{;\mu;\rho}
\]
\[
-16 R^\mu_\nu\Phi_{;\mu;\rho}\Phi^{;\nu;\rho}
-2R_{\mu\nu}R^{\mu\nu}\Phi_{;\rho}\Phi^{;\rho}
\]
\[
+4R_{\mu\nu}R^{\mu\nu}\Phi\Phi^{;\rho}_{;\rho}
+\frac{5}{3}R_{\mu\nu}R^{\mu\nu}R\Phi^2
\]
\[
-8R^{\mu\nu}\Phi^{;\rho}_{;\nu}\Phi_{;\rho;\mu}
-2\Phi^2R_{\mu\nu}R^{\nu\rho}R_\rho^\mu~,
\]
\[
S_3= -\frac{8}{3\Phi}R\Phi_{;\mu;\nu}\Phi^{;\mu}\Phi^{;\nu}
\]
\[
+\frac{8}{3\Phi} R \Phi^{;\mu}\Phi_{;\mu}\Phi_{;\nu}^{;\nu}
-\frac{8}{\Phi} R_{\mu\nu}\Phi^{;\mu;\nu}\Phi_{;\rho}\Phi^{;\rho}~.
\]
Here $;$ as usual denotes a covariant derivative. This action is invariant under general coordinate and Weyl transformations by construction. On the flat background, the actions $S_2$ and $S_3$ are equal to zero. One can check that after a few integrations by parts the action $S_1$ is nothing but ${\cal O}_1$ if the metric is flat.

\bibliography{PRD_misha}
\end{document}